\documentclass[12pt]{iopart}
\usepackage{graphicx}
\usepackage{hyperref}

\hypersetup{pdfborder={0 0 0}}

\makeatletter
\renewcommand\section{\@startsection {section}{1}{\z@}
  {3.0ex minus 1.0ex}
  {2.0ex}
  {\reset@font\normalsize\bfseries\raggedright}
}
\renewcommand\subsection{\@startsection{subsection}{2}{\z@}
  {3.0ex minus 1.0ex}
  {1.0ex}
  {\reset@font\normalsize\bfseries\raggedright}
}
\renewcommand\subsubsection{\@startsection{subsubsection}{3}{\z@}
  {2.0ex minus 1.0ex}
  {0.5ex}
  {\reset@font\normalsize\itshape}
}
\makeatother

\begin{document}

\sloppy
\flushbottom

\twocolumn[

    \title{SpikeSift: A Computationally Efficient and Drift-Resilient Spike Sorting Algorithm}
        
    \author{V Georgiadis and P C Petrantonakis}
        
    \address{Department of Electrical and Computer Engineering, Aristotle University of Thessaloniki, GR}
        
    \eads{\mailto{georgiadv@ece.auth.gr}, \mailto{ppetrant@ece.auth.gr}}
        
    \begin{abstract}
        Objective: Spike sorting is a fundamental step in analysing extracellular recordings, enabling the isolation of single‑neuron activity. However, it remains a challenging problem because extracellular traces mix overlapping spikes from neighbouring cells and are marred by recording instabilities such as electrode drift. Numerous algorithms have been proposed, yet many struggle to balance accuracy and computational efficiency, limiting their practicality for today’s large‑scale datasets.
        Approach: In response, we introduce SpikeSift, a spike‑sorting algorithm expressly designed to mitigate drift while running on standard CPUs. SpikeSift (i) partitions long recordings into shorter, relatively stationary segments, (ii) carries out spike detection and clustering simultaneously through an iterative detect‑and‑subtract scheme within each segment, and (iii) preserves neuronal identity across segments via a fast template‑alignment stage that dispenses with continuous trajectory estimation.
        Main results: Extensive validation on paired intracellularly validated datasets and on biophysically realistic MEArec simulations---covering elevated noise, diverse drift profiles, ultra‑short recordings and bursting activity---demonstrates that SpikeSift matches or exceeds the accuracy of state‑of‑the‑art methods while completing sorting an order of magnitude faster on a single desktop core.
        Significance: The combination of high fidelity, drift resilience, and modest computational demand renders SpikeSift broadly accessible while preserving data quality for downstream neurophysiological analysis.
    \end{abstract}
        
    \noindent
    {\it Keywords: Spike sorting, Electrode drift, Template matching}
    \vspace{2.5ex}
]

\section{Introduction}
\label{introduction}

\noindent 
Neuronal communication occurs through action potentials---brief electrical impulses that encode and transmit information across neural circuits \cite{action_potentials}. Analyzing these signals is essential for understanding brain function \cite{encoding_information}, diagnosing neurological disorders \cite{neurological_disorder}, and advancing neurotechnologies, including brain-computer interfaces \cite{brain_computer_interfaces}. While intracellular recordings provide highly precise measurements by inserting microelectrodes directly into neurons \cite{patch_clamp}, their invasive nature limits their feasibility for large-scale and long-term studies \cite{large_scale_recordings}. Extracellular recordings offer a less invasive alternative by capturing voltage fluctuations in the surrounding medium; yet the resulting voltage traces represent the composite activity of nearby cells \cite{extracellular_fields}. Work spanning more than two decades has shown that useful information can sometimes be extracted directly from this composite signal, without assigning spikes to individual neurons.  Decoders that operate on threshold-crossing multi-unit activity or local-field potentials have reconstructed limb kinematics in motor cortex \cite{lfp_decoding}, and Bayesian models have decoded an animal’s spatial position from unsorted hippocampal activity \cite{unsorted}.  Spectral--temporal embeddings of raw extracellular signals can likewise capture task-related dynamics without explicit clustering \cite{spectral_representation}.  While such findings highlight the value of unsorted analyses, studies that probe synaptic connectivity, detailed population codes, or single-neuron biomarkers still demand cell-specific spike trains.  In these cases the composite signal must be disambiguated through spike sorting, the computational attribution of each detected spike to its most probable source neuron \cite{spike_sorting_overview}.

Spike sorting typically consists of four key steps: (1) bandpass filtering to isolate relevant frequency bands, (2) spike detection to identify candidate events, (3) feature extraction to reduce waveform dimensionality, and (4) clustering to assign spikes to putative neurons \cite{spike_sorting_review,superparamagnetic}. Over the years, a range of algorithms have been developed to optimize this process \cite{past_present_future}, each offering distinct advantages and limitations. Density-based clustering techniques, exemplified by ISO-SPLIT \cite{iso_split}, introduced in Mountainsort \cite{mountainsort}, allow flexible neuron identification but assume unimodal waveform distributions, an assumption that is not always valid in practice. Graph-based clustering methods, as implemented in Kilosort \cite{kilosort}, enhance spike separability by leveraging the spatial structure of detected spikes, albeit at the cost of increased computational complexity. Template-matching algorithms \cite{template_matching}, used in approaches like Kilosort and SpyKING CIRCUS \cite{circus}, achieve high precision by comparing detected spikes against predefined waveform templates, yet their computational demands remain substantial. More recently, deep learning models have been proposed to automate feature extraction and classification \cite{yass,removing_noise,autoencoders,deep_learning}, but their dependence on large labeled datasets poses a significant challenge, particularly in recordings with diverse neuronal activity.

Meanwhile, the advent of high-density multi-electrode arrays, exemplified by Neuropixels 2.0 \cite{neuropixels1,multielectrode_array,neuropixels2}, has dramatically increased the volume of neural data, amplifying the computational burden of spike sorting. While GPUs and FPGAs can accelerate processing \cite{hardware_implementations}, such hardware is not always available, rendering computational efficiency a central concern. A further obstacle to reliable spike sorting is electrode drift, where shifts in electrode positions disrupt waveform consistency \cite{neuropixels2}, leading to spurious assignments and cluster fragmentation. Approaches to counteracting drift generally fall into two broad categories. The first, global drift correction \cite{kilosort,drift_correction}, estimates the drift trajectory and applies interpolation-based realignment to preserve waveform integrity. While effective in some cases, residual inaccuracies can still result in fragmented clusters. The second, segmentation-based sorting \cite{mountainsort}, divides recordings into short time intervals under the assumption that drift is minimal within each segment. However, this approach depends on accurate spike sorting within short windows to prevent drift-induced waveform variations from degrading clustering performance.

Given these challenges, benchmark datasets with validated spike identities are essential for a comprehensive assessment of spike sorting accuracy \cite{spikeinterface,spikeforest}. The most reliable method involves simultaneous intracellular and extracellular recordings \cite{circus}, enabling direct ground-truth comparisons. However, this approach remains technically demanding and feasible for only a limited number of neurons. To complement these experimental benchmarks, biophysically realistic simulations provide a scalable framework for evaluating sorting accuracy in scenarios that are difficult to reproduce in vivo \cite{neuron,visapy,mearec}.

\begin{figure*}[t]
    \centering
    \includegraphics[width=0.9\textwidth,height=0.48\textwidth]{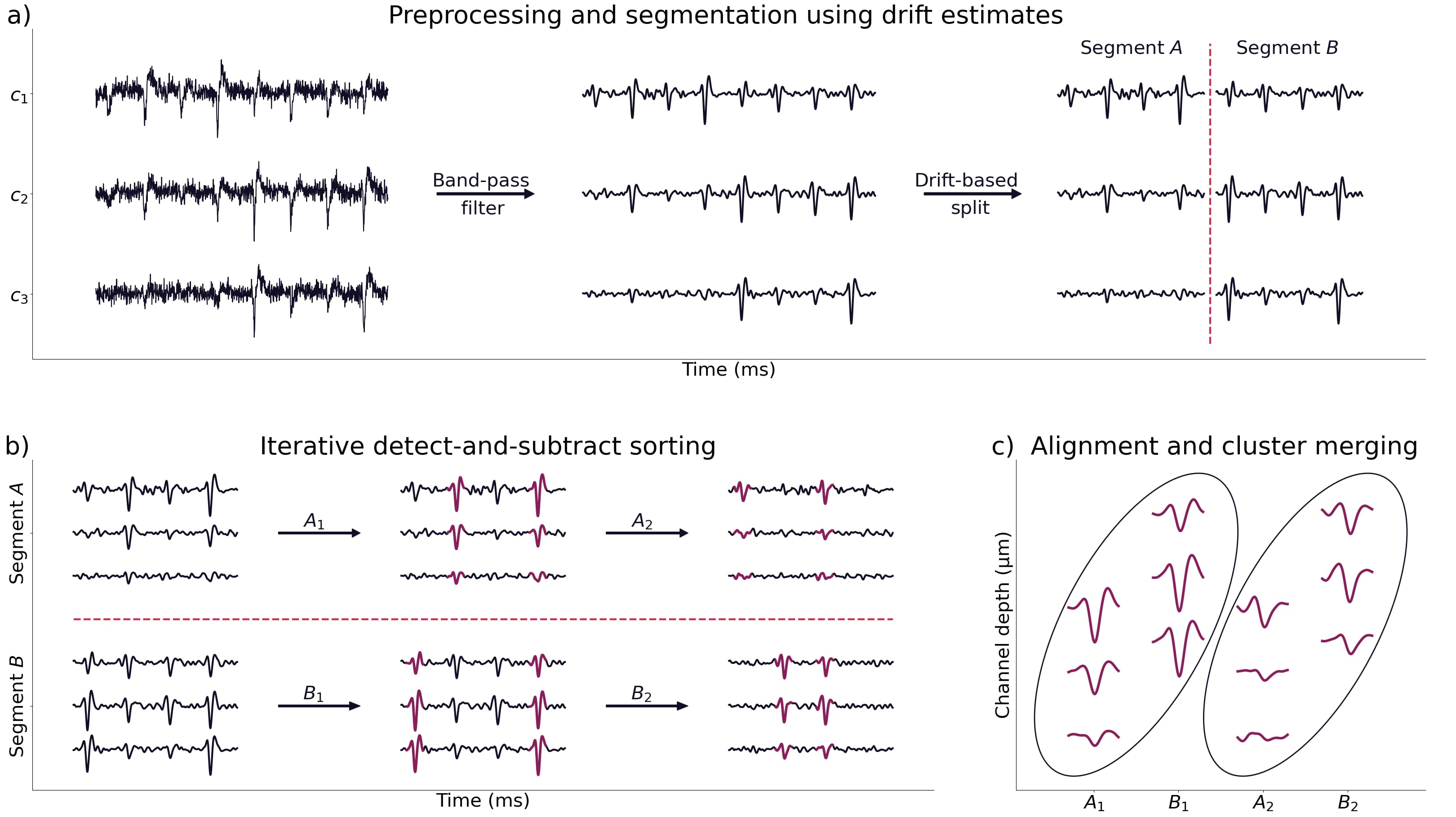}
    \caption{Overview of SpikeSift's three-stage workflow.
    (a) A raw extracellular signal recorded from three channels, displayed in three traces from left to right: the unfiltered signal, the bandpass-filtered signal (isolating spike-relevant frequencies), and the segmented signal. The recording is divided into two intervals (Segment A and Segment B) at the dashed line, where amplitude fluctuations suggest electrode drift. Specifically, beyond this point, spikes on channels $c_2$ and  $c_3$ increase in amplitude, while those on $c_1$ decrease, indicating a relative shift in neuron positions.
    (b) Within each segment, spikes are detected and subtracted iteratively, allowing neurons to be identified sequentially. Once a unit is detected, its average waveform is subtracted from the recording, unveiling additional spikes in subsequent iterations. Two neuronal units per segment ($A_1$, $A_2$ in Segment $A$ and $B_1$, $B_2$ in Segment $B$) are illustrated.
    (c) Localized drift compensation ensures that neuronal clusters remain consistently identified across segment boundaries. The average waveforms of two neurons ($A_1$, $A_2$) from Segment $A$ are matched to their counterparts ($B_1$, $B_2$) in Segment $B$. A relative spatial shift is applied to align corresponding units, preserving neuronal identity across segments.
    }
    \label{fig:pipeline}
\end{figure*}

To address these challenges, we introduce SpikeSift, a computationally efficient spike sorting algorithm designed to ensure sorting accuracy even in the presence of electrode drift. As illustrated in Figure~\ref{fig:pipeline}, SpikeSift employs a three-stage workflow: (1) adaptive segmentation, which divides the recording into segments where drift is less pronounced, (2) integrated spike detection and clustering within an iterative detect-and-subtract framework, and (3) cluster alignment across segments to ensure consistent neuronal identities throughout the recording. By leveraging a lightweight template-matching process, SpikeSift enables rapid processing on standard hardware, making it well-suited for large-scale neuroscience studies and exploratory research, where both accuracy and computational feasibility are critical.

\section{Methods}
\label{methods}

\noindent 
SpikeSift enhances clustering reliability by segmenting the recording into shorter intervals where neuronal waveforms are less affected by drift. Although segmentation-based approaches have been explored in prior work \cite{mountainsort}, they typically rely on fixed, user-defined time windows that may fail to capture abrupt shifts when drift occurs unpredictably. By contrast, SpikeSift determines segment boundaries dynamically, as illustrated in Figure~\ref{fig:pipeline}a, analyzing fluctuations in spike amplitudes to identify intervals where drift-induced variations are minimal. This adaptive strategy ensures that neuronal activity remains coherent within each segment, providing a robust foundation for subsequent sorting. A detailed description of this process is provided in Section~\ref{sec:phase1}.

Within these segments, SpikeSift employs an iterative detect-and-subtract approach, as illustrated in Figure~\ref{fig:pipeline}b, to balance detection sensitivity, clustering accuracy, and computational efficiency. A fundamental limitation of conventional pipelines is that spike detection---whether based on amplitude thresholds or template matching---is performed separately from clustering \cite{past_present_future}, meaning that clustering operates on all detected spikes without inherently filtering out spurious detections. This structure imposes a trade-off: a strict detection threshold minimizes false positives but may lead to incomplete clusters, where detected neurons are missing a significant portion of their spikes. Conversely, a more permissive threshold captures additional spikes but increases the risk of spurious detections, which can give rise to artificial neuronal units. 

To address this limitation, SpikeSift integrates detection and clustering into an adaptive, multi-stage framework. The process begins with a strict detection threshold that isolates only the most prominent spikes, ensuring the formation of a well-defined template by identifying a single coherent cluster---even if some spikes remain undetected. This template is then used in a lightweight matching step to selectively recover additional spikes with similar waveforms, increasing detection sensitivity while tolerating some false positives. However, spurious detections introduced at this stage are systematically removed in the final clustering step, which retains only spikes that conform to the established template. This adaptive filtering process dynamically refines the cluster boundaries, ensuring that neuronal units remain well-defined and complete. Once a neuronal unit is fully resolved, its spikes are subtracted from the recording, and the process repeats iteratively until no further spikes remain. By embedding clustering within an adaptive detection loop, SpikeSift overcomes the limitations of threshold-based sorting, achieving both high accuracy and computational efficiency, as detailed in Section~\ref{sec:phase2}.

Once within-segment sorting is complete, SpikeSift aligns clusters across segment boundaries to preserve neuronal identities throughout the recording, as illustrated in Figure~\ref{fig:pipeline}c. Conventional drift correction methods \cite{kilosort,drift_correction} attempt to estimate a continuous drift trajectory and apply waveform realignment transformations, but these approaches often introduce residual inaccuracies, leading to cluster fragmentation. In contrast, SpikeSift leverages the stability of neuronal clusters within each segment, assuming that a neuron remains consistently represented by a single cluster. Instead of relying on strict waveform realignment, it establishes correspondences between clusters based on their amplitude signatures across channels, ensuring a coherent mapping of neuronal identities across segments. A lightweight interpolation model accounts for waveform shifts, improving accuracy without imposing excessive computational costs. By prioritizing cluster continuity over waveform realignment, SpikeSift maintains neuronal identities while reducing the fragmentation errors commonly associated with conventional drift correction methods, as detailed in Section~\ref{sec:phase3}.

\begin{figure*}[t]
    \centering
    \includegraphics[width=0.9\textwidth,height=0.54\textwidth]{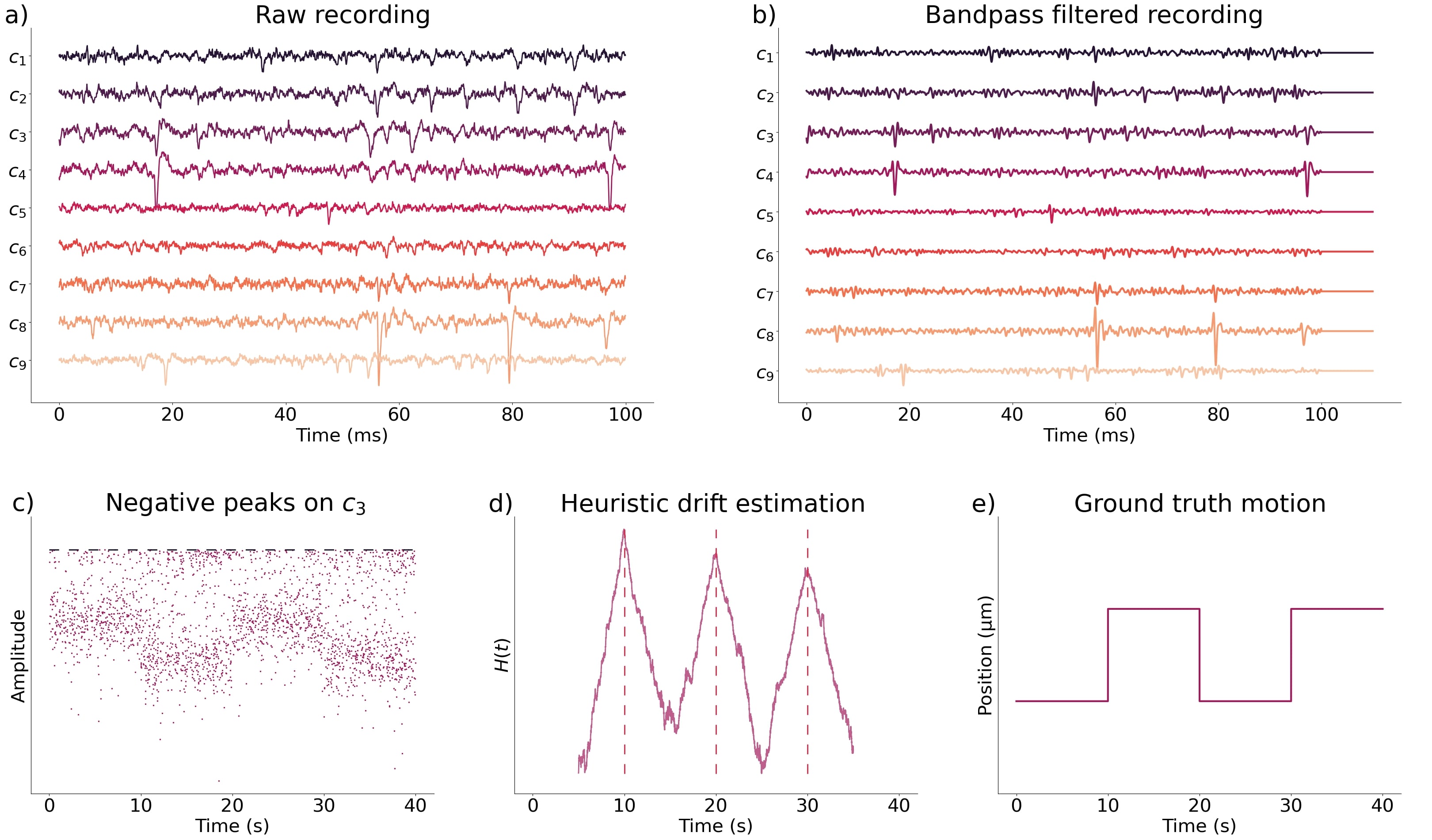}
    \caption{Filtering and adaptive segmentation.
    (a) A short segment of raw extracellular data, with each channel represented in a different color.
    (b) The same trace after applying a Difference-of-Gaussians ($DoG$) filter, which selectively enhances frequencies in the 300--$3000 Hz$ range while suppressing low-frequency fluctuations and excessive high-frequency noise.
    (c) Negative peaks on channel $c_3$, with the dashed line representing the dynamic detection threshold $\theta$. Systematic shifts in peak amplitudes suggest potential electrode drift.
    (d) The heuristic drift measure $H(t)$, computed from amplitude fluctuations across all channels. Peaks in $H(t)$ (dashed lines) indicate time points where amplitude fluctuations are most pronounced, enabling dynamic segmentation.
    (e) The ground truth motion of the recording, confirming that the detected segmentation points correspond to moments of pronounced drift.
    }
    \label{fig:phase1}
\end{figure*}

\subsection{Filtering and Segmentation}
\label{sec:phase1}

\noindent 
To ensure robust spike sorting, SpikeSift preprocesses extracellular recordings in three key stages. First, bandpass filtering (Section~\ref{subsec:filtering}) enhances spike waveforms by attenuating low-frequency fluctuations and high-frequency noise. Next, spike detection (Section~\ref{subsec:detection}) identifies candidate spikes, providing reference points for tracking amplitude variations over time. Finally, adaptive segmentation (Section~\ref{subsec:segmentation}) divides the recording at points where drift-induced fluctuations are most pronounced, improving clustering accuracy without requiring explicit trajectory estimation.

\subsubsection{Difference-of-Gaussians}
\label{subsec:filtering}

\noindent 
Extracellular recordings contain a mix of signals, including low-frequency local field potentials and high-frequency noise \cite{extracellular_fields}, both of which can obscure spike waveforms, as shown in Figure~\ref{fig:phase1}a. To isolate the 300--$3000 Hz$ frequency band commonly used for spike detection \cite{past_present_future,deep_learning}, SpikeSift applies a Difference-of-Gaussians ($DoG$) filter---a computationally efficient bandpass approximation \cite{gaussian_filter} that enhances spike visibility while suppressing irrelevant components. As shown in Figure~\ref{fig:phase1}b, this filtering step effectively attenuates low-frequency fluctuations and high-frequency noise, preserving the integrity of neuronal spikes.

In most spike sorting pipelines, filtering is a relatively minor computational step. However, due to SpikeSift’s efficiency, the relative processing burden shifts, making filtering a major contributor to overall runtime. The $DoG$ filter was chosen for its ability to run more than twice as fast as conventional bandpass filters while preserving signal fidelity. Despite this speed advantage, filtering still accounts for nearly one-third of total processing time, making it a key bottleneck and highlighting the need for further optimizations to enhance overall efficiency.

Mathematically, the $DoG$ filter approximates a Gaussian filter with standard deviation $\sigma$ using a cascade of four box filters:

\vspace{2ex}
\centerline{$ G_{\sigma}[n] \approx (U_W * U_W * U_W * U_W)[n] $}
\vspace{2ex}

\noindent 
where $W \approx \sigma \sqrt{3}$. The bandpass effect is then achieved by subtracting two such filters with different standard deviations:

\vspace{2ex}
\centerline{$ DoG[n] \approx (U_{W_1}^4 - U_{W_2}^4)[n] $}
\vspace{2ex}

\noindent 
where $W_1$ and $W_2$ are selected to approximate the 300--$3000 Hz$ frequency range.

The resulting band-passed trace is retained for both detection and clustering. Although band-pass filters can distort fine waveform shape and phase, such effects mainly hinder shape-based sorters \cite{unfiltered,filter_distortions}. SpikeSift instead distinguishes units through the relative peak amplitudes recorded on neighbouring electrodes (see Section \ref{subsec:clustering}). Pilot checks showed that re-extracting snippets from the unfiltered signal increased noise in those amplitude ratios and produced no measurable gain in unit separability. Within this amplitude-centred framework, the filtered trace offers a practical balance between signal clarity and computational cost.

\subsubsection{Spike Detection}
\label{subsec:detection}

\noindent 
After filtering, SpikeSift detects spikes as negative voltage deflections that cross a dynamic threshold $\theta$ (dashed line in Figure~\ref{fig:phase1}c). This preference for negative peaks stems from the biophysical properties of extracellular action potentials, which typically manifest as negative deflections due to the outward flow of positive ions \cite{extracellular_fields,negative_spikes}. Although SpikeSift is optimized for detecting negative peaks, it can be seamlessly adapted for positive deflections by inverting the sign of the $DoG$ filter, without incurring additional computational overhead.

To ensure a robust and adaptive detection threshold, SpikeSift employs the median absolute deviation (MAD) method \cite{superparamagnetic,past_present_future}, a widely used approach in neural recordings due to its resilience to outliers and its effectiveness in handling non-Gaussian noise distributions. The threshold is defined as:

\vspace{2ex}
\centerline{$ \theta = -\kappa \cdot \mathrm{MAD} $}
\vspace{2ex}

\noindent 
where $\kappa$ is a user-defined parameter that controls detection sensitivity. Larger values limit detection to only the most prominent spikes, whereas smaller values capture additional low-amplitude waveforms at the cost of increased computational complexity.

\subsubsection{Adaptive Segmentation}
\label{subsec:segmentation}

\noindent 
Detected spikes serve as reference points for tracking waveform variations over time. Rather than relying on fixed time intervals, SpikeSift dynamically determines segment boundaries based on amplitude fluctuations, minimizing within-segment drift and preserving clustering accuracy.

Traditional drift correction methods often attempt to estimate neuron positions directly using monopolar triangulation \cite{drift_correction,monopolar}, which models neurons as point sources emitting extracellular fields with symmetrical propagation. However, this assumption does not always hold in complex neural environments, where heterogeneous tissue properties and overlapping sources distort signals \cite{extracellular_fields}.

SpikeSift takes an alternative approach, inferring drift from fluctuations in spike amplitudes, which serve as a proxy for changes in neuron-to-electrode distance \cite{neuropixels2}. By identifying time points where these fluctuations are most pronounced, it adaptively partitions the recording to minimize drift-induced distortions without requiring explicit trajectory estimation.

To ensure a balance between drift adaptation and sorting reliability, SpikeSift enforces a minimum segment duration, $L_{\min}$, preventing segments from becoming too short or excessively long. Shorter segments improve drift compensation by capturing fine-grained motion, whereas longer segments enhance clustering stability, particularly for neurons with low firing rates.

To determine segmentation points, SpikeSift computes the summed spike amplitude over a sliding window:

\vspace{2ex}
\centerline{$ S(t) = \sum_{p \in (t, t+L_{\min})} (A_p - \theta) $}
\vspace{2ex}

\noindent 
where $A_p$ represents the amplitude of peak $p$ and subtracting $\theta$ reduces sensitivity to fluctuations near the detection threshold. 

To determine where drift is most pronounced, a drift-sensitive measure $H(t)$ is computed across all channels:

\vspace{2ex}
\centerline{$ H(t) = \sum_{channels} \left| S(t) - S(t - L_{\min}) \right| $}
\vspace{2ex}

\noindent
Because the summed amplitude $S(t)$ is computed relative to the detection threshold, $H(t)$ captures changes in each unit’s excursion above baseline rather than absolute spike height. Moderate-amplitude neurons therefore influence boundary placement alongside larger ones, and the segmentation adapts automatically if the most prominent units diminish or drop out during a recording.

Peaks in $H(t)$---marked by the dashed lines in Figure~\ref{fig:phase1}d---indicate periods of pronounced amplitude fluctuations across channels, corresponding to moments of significant electrode drift. As shown in Figure~\ref{fig:phase1}e, the detected boundaries closely align with actual electrode motion, demonstrating SpikeSift’s ability to accurately identify sharp drift events without relying on explicit trajectory estimation.

\begin{figure*}[t]
    \centering
    \includegraphics[width=0.9\textwidth,height=0.51\textwidth]{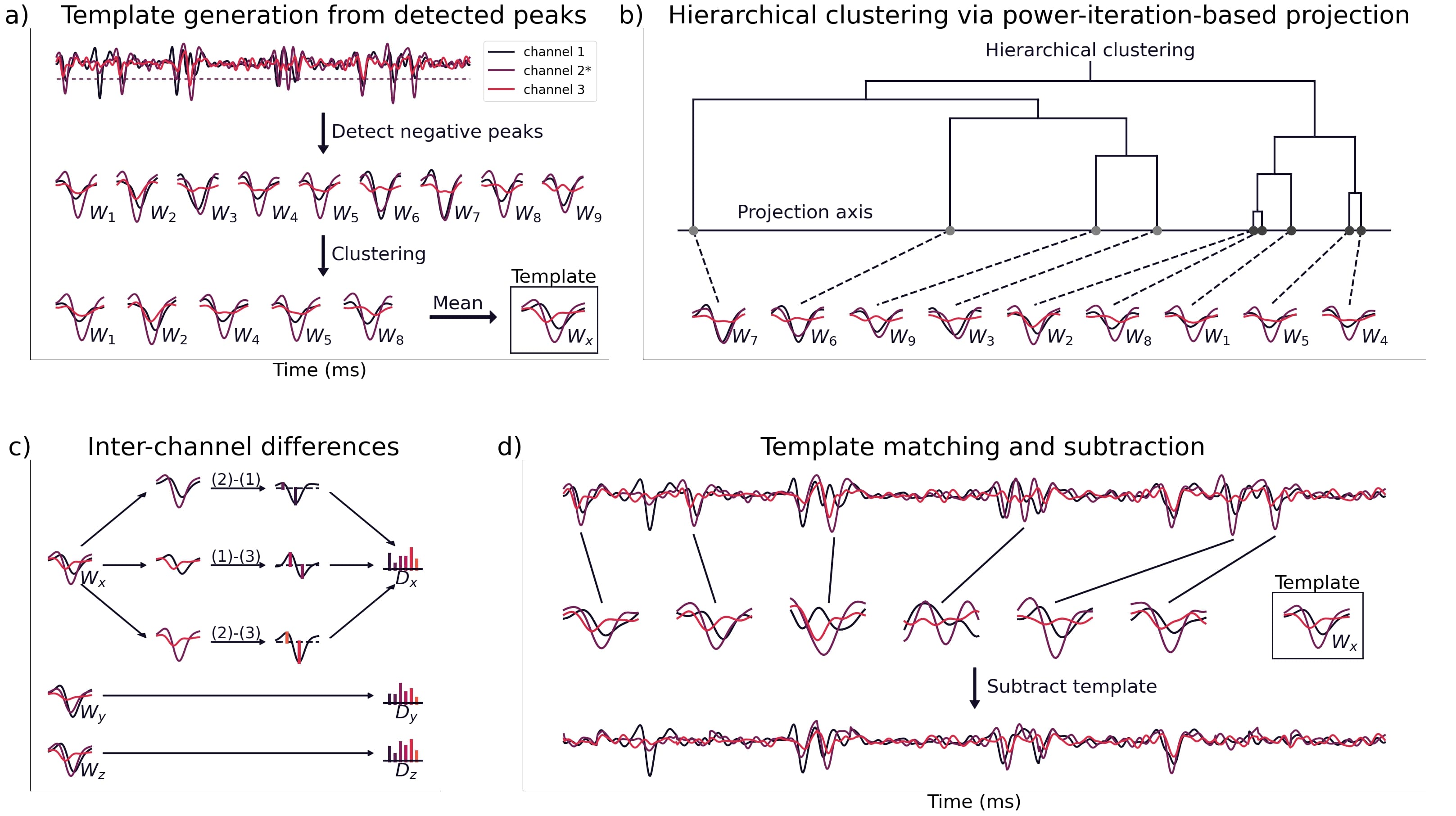}
    \caption{Within-segment spike sorting using an iterative detect-and-subtract approach.
    (a) A short recording from three channels (visualized for clarity, though SpikeSift operates on five channels). The algorithm selects the channel with the strongest negative peaks (Channel 2 in this example) and detects potential spike events when the signal crosses an adaptive threshold $\theta$ (red dashed line). Nine extracted waveforms ($W_1$--$W_9$) are shown, which are subsequently clustered using a binary-splitting procedure. The average waveform of the retained cluster ($W_1$,$W_2$,$W_4$,$W_5$,$W_8$) forms the template waveform $W_x$.
    (b) One step in the binary-splitting process. Waveforms are projected onto their principal axis of variance, followed by hierarchical clustering, which separates them into two candidate clusters (light vs. dark points).
    (c) Inter-channel difference vectors are used to determine whether candidate clusters originate from distinct neurons. These vectors are visualized as colored bars, where each color represents a different electrode pair. Three clusters ($x$,$y$,$z$) are compared: although $W_z$ has a similar amplitude to $W_y$, its inter-channel difference vector $D_z$ more closely resembles $D_x$. The ground truth confirms that spikes from $x$ and $z$ belong to the same neuron, demonstrating that spatial waveform structure provides a more reliable way to distinguish neurons than amplitude alone.
    (d) Template subtraction: All spikes matching the identified template are detected and subtracted from the recording, reducing interference in subsequent iterations. This allows lower-amplitude or overlapping spikes to be detected in later passes, improving sorting accuracy.
    }
    \label{fig:phase2}
\end{figure*}

\subsection{Within-Segment Iterative Sorting}
\label{sec:phase2}

\noindent 
Following segmentation, SpikeSift employs an iterative detect-and-subtract strategy to systematically identify neurons within each segment. Unlike conventional pipelines, which treat spike detection and clustering as independent processes, this approach integrates them into a unified framework that incrementally isolates individual neurons. The sorting procedure alternates between two primary phases: first, a binary-splitting clustering strategy (Section~\ref{subsec:clustering}) is employed to identify a single neuron, forming a representative template (Section~\ref{subsec:formation}). Then, this template is used to detect and eliminate all corresponding spike occurrences through template matching (Section~\ref{subsec:matching}). By restricting detection to spikes that closely resemble the current template, this approach not only enhances clustering accuracy but also significantly improves computational efficiency.

\subsubsection{Template Formation}
\label{subsec:formation}

\noindent 
To construct a neuronal template, SpikeSift first selects a reference channel, $c^*$, based on cumulative spike amplitude, prioritizing channels with consistently strong signals over those dominated by a single large outlier. The reference channel is determined using the criterion:

\vspace{2ex}
\centerline{$ c^*=\arg\max_c \sum (A_p - \theta) $}
\vspace{2ex}

\noindent 
where $A_p$ represents the amplitude of peak $p$, and $\theta$ is the detection threshold defined in Section~\ref{subsec:detection}. By selecting the channel with the consistently strongest spike activity, this approach ensures that template formation focuses on neurons with reliable signals, thereby improving detection robustness.

Once $c^*$ is designated, spike detection proceeds as outlined in Section~\ref{subsec:detection}. For each detected spike, a 2-ms waveform \cite{spike_sorting_overview,past_present_future} is extracted from the five electrodes closest to $c^*$, capturing the spike's spatial distribution while maintaining computational efficiency \cite{limited_electrodes}, as shown in Figure~\ref{fig:phase2}a:

\vspace{3ex}
\centerline{$ W_i=\{s_c^{(t_i\pm 1ms)} \mid c \in \mathcal{N}(c^*,5)\} $}
\vspace{3ex}

\noindent 
where $s_c^{(t)}$ represents the recorded signal on channel $c$ at time $t$, and $\mathcal{N}(c^*,5)$ represents the set of the five electrodes nearest to $c^*$.

These waveforms then undergo a clustering step, as described in Section~\ref{subsec:clustering}, to isolate the activity of a single neuronal unit. The mean waveform of the resulting cluster---illustrated in Figure~\ref{fig:phase2}a---serves as a template for subsequent spike identification in Section~\ref{subsec:matching}, and is computed as:

\vspace{3ex}
\centerline{$ W_x=\frac{1}{|C_x|}\sum_{W_i \in C_x}W_i $}
\vspace{3ex}

\noindent 
where $C_x$ represents the set of waveforms assigned to the extracted neuronal unit $x$.

\subsubsection{Binary-Splitting Clustering}
\label{subsec:clustering}

\noindent 
SpikeSift’s clustering process follows an iterative refinement strategy, progressively filtering out outliers until a single, well-defined cluster remains. The algorithm first projects waveforms onto their principal axis of variance---illustrated in Figure~\ref{fig:phase2}b---using an approximation via the power iteration method \cite{power_iteration}:

\vspace{3ex}
\centerline{$ v^* \approx \arg\max_v\frac{v^TCv}{v^Tv} $}
\vspace{3ex}

\noindent 
where $C$ represents the covariance matrix of the waveforms. Since each iteration performs only a single binary split, this projection provides sufficient separation while maintaining computational efficiency.

Next, hierarchical clustering is performed using an optimized nearest-neighbor-chain algorithm \cite{hierarchical_clustering}, reducing worst-case complexity from $O(n^3)$ to $O(n^2)$, with additional optimizations leveraging the one-dimensional nature of the data to achieve $O(n \log n)$ complexity. The hierarchical clustering tree---visualized in Figure~\ref{fig:phase2}b---illustrates the progressive merging of clusters based on the following metric:

\vspace{3ex}
\centerline{$ \|m_{x}-m_{y}\|^2 \cdot \min(|C_x|,|C_y|) $}
\vspace{3ex}

\noindent 
where $m_{x}$ and $m_{y}$ denote the mean values of the projected waveforms in clusters $C_x$ and $C_y$. This formulation prioritizes merging smaller clusters first, preventing the premature fusion of well-separated neuronal units.

Once only two clusters remain, SpikeSift determines whether they originate from the same neuron by assessing waveform differences. If they are deemed to belong to the same neuron, they are merged, concluding the clustering process. Otherwise, the cluster with the larger average spike amplitude is further subdivided, while lower-amplitude signals are deferred to subsequent iterations.

Because extracellular spikes typically exhibit large negative amplitudes \cite{extracellular_fields,negative_spikes}, direct waveform comparisons can be misleading, as relative differences can appear small in proportion to overall magnitude. To mitigate this issue, SpikeSift evaluates inter-channel voltage differences---illustrated in Figure~\ref{fig:phase2}c---capturing spatial waveform structure more effectively. This approach enhances neuronal separation while eliminating the need for computationally expensive whitening, which is commonly used to decorrelate channels \cite{kilosort}.

To derive the inter-channel difference vector $D_x$ for a given cluster $x$, SpikeSift first computes the mean waveform $W_x$ as described in Section~\ref{subsec:formation}. Then, for each ordered pair of recording channels $(c_i,c_j)$, it determines the maximum amplitude difference within the waveform window:

\vspace{3ex}
\centerline{$ D_x^{(c_i,c_j)}=\max(W_x^{(c_i)}-W_x^{(c_j)}) $}
\vspace{3ex}

\noindent 
This difference vector $D_x$ serves as a compact representation of the waveform's spatial structure across electrodes, capturing how the neuron's extracellular signature varies across the array.

The two clusters are merged if their Euclidean distance in this difference space falls below a predefined threshold $\lambda$:

\vspace{3ex}
\centerline{$ \|D_x - D_y\| \leq \lambda \cdot \max(\|D_x\|, \|D_y\|) $}
\vspace{3ex}

\noindent 
This formulation accounts for inherent signal variability while ensuring that activity from distinct neurons remains well separated. By leveraging spatial waveform structure rather than absolute amplitude differences, it effectively discriminates between neurons while preventing excessive fragmentation.

\subsubsection{Template Matching and Subtraction}
\label{subsec:matching}

\noindent 
Once a template waveform is formed, SpikeSift identifies all spikes that conform to the template within the segment. Candidate spikes are detected as local minima on the reference channel $c^*$, where a local minimum is defined as a peak that is neither preceded nor followed by a larger peak within $1 ms$. This criterion ensures that closely spaced spikes, such as those occurring during bursting activity \cite{burst_firing}, are accurately captured while avoiding redundant detections.

Given the high spike detection rate---ranging from 500 to 1000 spikes per second---relative to typical neuronal firing rates, which vary from less than one to several dozen spikes per second \cite{firing_rates}, an initial filtering step is applied to reduce computational overhead before clustering. For each candidate spike, the algorithm constructs a five-channel feature vector:

\vspace{3ex}
\centerline{$ v_i=\{s_c^{(t_i)} \mid c \in \mathcal{N}(c^*,5)\} $}
\vspace{3ex}

\noindent 
where $s_c^{(t)}$ represents the recorded signal on channel $c$ at time $t$, and $\mathcal{N}(c^*,5)$ denotes the set of the five channels closest to $c^*$. Candidates that are closer to the origin than to the template vector $T$ are discarded using the criterion:

\vspace{2ex}
\centerline{$ v_i \cdot T < \frac{1}{2}\|T\|^2 $}
\vspace{2ex}

\noindent 
Since the origin represents the absence of spikes, this filtering step effectively eliminates noise while preserving relevant neuronal activity. The binary-splitting clustering method from Section~\ref{subsec:clustering} is then applied. However, instead of selecting the cluster with the largest amplitude, the algorithm retains the one that best resembles the template.

To ensure clustering reliability, only clusters containing at least $N_{\min}$ spikes are retained. This introduces a fundamental trade-off: reducing $N_{\min}$ enables the detection of low-firing neurons but increases the risk of forming clusters with insufficient statistical support. Similarly, clusters with an average waveform amplitude below the detection threshold $\theta$ are discarded, as these spikes are unlikely to be distinguishable from background noise---particularly given that the template was derived from spikes exceeding this threshold.

If a cluster is discarded, the channel from which its template was extracted is excluded from serving as a reference channel in subsequent iterations. Since the algorithm has already failed to identify a stable cluster from that channel, further attempts are unlikely to succeed, making continued analysis redundant. Removing such channels ensures that computational resources are concentrated on those more likely to yield reliable spike templates.

This process repeats until no valid reference channels remain, at which point the algorithm terminates. After each iteration, the average waveform of the identified spikes is subtracted from the segment \cite{hardware_implementations}, progressively revealing lower-amplitude spikes that may have been masked by stronger signals in earlier iterations---an effect illustrated in Figure~\ref{fig:phase2}d.

\begin{figure*}[t]
    \centering
    \includegraphics[width=0.9\textwidth,height=0.50\textwidth]{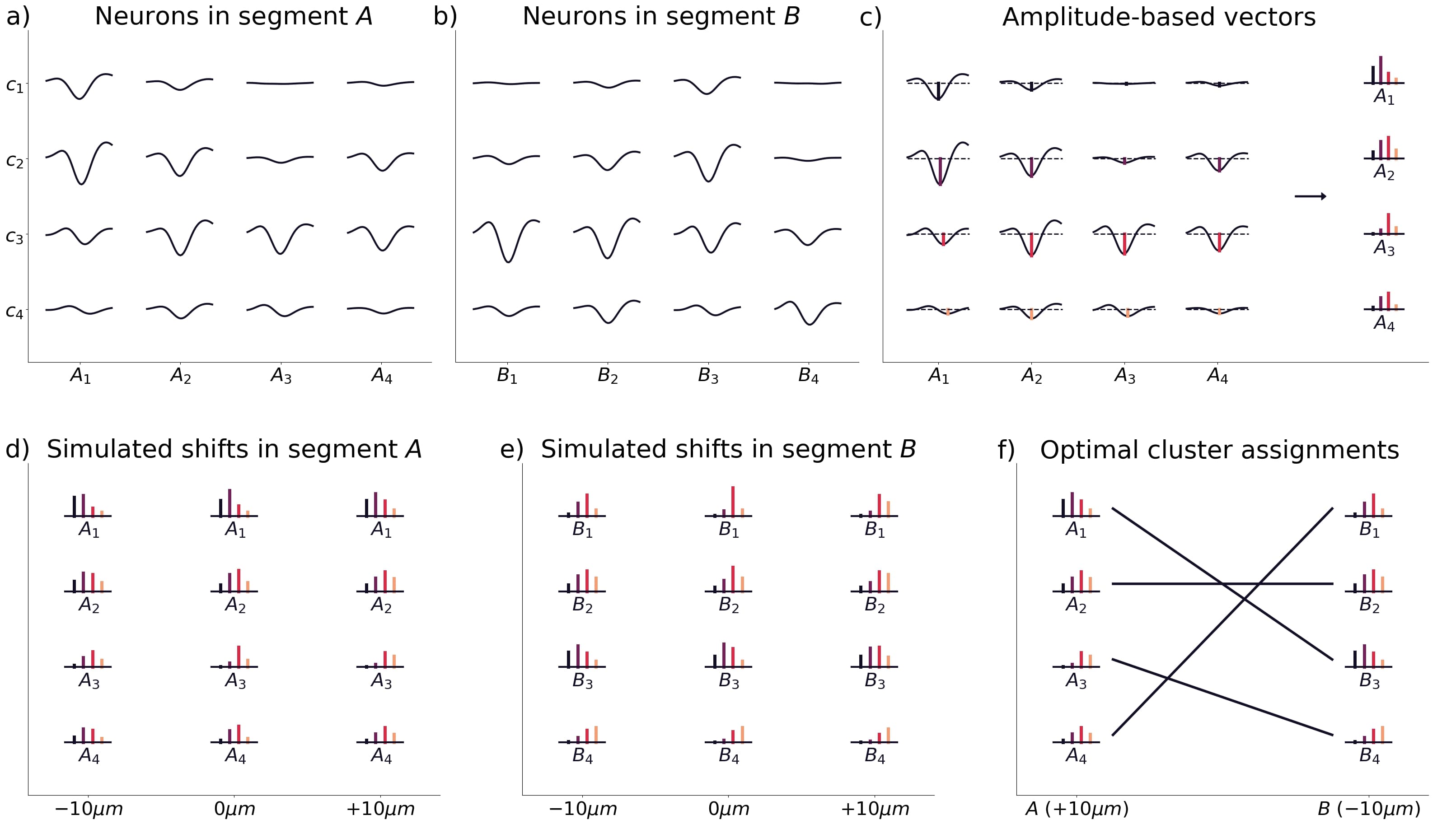}
    \caption{Localized alignment of clusters across segments.
    (a) Average waveforms of four neurons ($A_1$--$A_4$) in Segment $A$ recorded across four channels.
    (b) Corresponding average waveforms of four neurons ($B_1$--$B_4$) in Segment $B$ recorded on the same channels. Differences in waveform amplitudes suggest potential electrode drift between segments.
    (c) Each cluster’s average waveform is represented as a channel-specific amplitude vector, visualized as colored bars. These bars illustrate waveform amplitudes across channels rather than representing a histogram.
    (d) Simulated electrode shifts ($\pm 10 \mu$m) applied to $A_1$--$A_4$. Linear interpolation estimates how amplitude vectors change with small displacements, modeling the effects of drift.
    (e) Equivalent simulated shifts applied to $B_1$--$B_4$.
    (f) Final cluster assignments after selecting the optimal shift for each segment. For example, $A_1$ aligns with $B_3$, $A_2$ with $B_2$, $A_3$ with $B_4$, and $A_4$ with $B_1$, ensuring that neuron identities remain consistent across segment boundaries.
    }
    \label{fig:phase3}
\end{figure*}

\subsection{Merging Clusters Across Segments}
\label{sec:phase3}

\noindent 
Although segmentation ensures that each neuron is consistently represented by a single cluster within each segment, waveform distortions may still arise at segment boundaries due to electrode displacement \cite{neuropixels2}, as illustrated in Figures~\ref{fig:phase3}a and~\ref{fig:phase3}b. Conventional global drift correction methods attempt to continuously track neuron positions, applying interpolation-based realignment to compensate for motion \cite{kilosort,drift_correction}. However, these approaches rely on highly precise spatial transformations, yet the inherently non-linear nature of electrode drift makes such accuracy difficult to achieve, often resulting in residual misalignments that fragment clusters.

In contrast, SpikeSift preserves neuronal identities by leveraging the natural one-to-one correspondence between clusters in consecutive segments. To accomplish this, the algorithm first encodes waveforms as compact amplitude vectors (Section~\ref{subsec:amplitude}), capturing essential waveform features while minimizing computational complexity. It then simulates electrode drift (Section~\ref{subsec:simulation}) by systematically testing possible displacements along the probe axis to identify the shift that best aligns amplitude vectors. Finally, cluster alignment is performed using a linear assignment framework (Section~\ref{subsec:assignment}), which establishes correspondences between neurons in adjacent segments.

\subsubsection{Amplitude-Based Representation}
\label{subsec:amplitude}

\noindent 
Instead of processing full waveforms, SpikeSift represents each neuron using a compact amplitude vector, significantly reducing computational overhead. As illustrated in Figure~\ref{fig:phase3}c, this transformation encodes each cluster $x$ solely by its maximum negative deflection on each channel, capturing essential waveform features while maintaining efficiency.

\vspace{3ex}
\centerline{$ A_x^{(c)}=\max_t -W_x^{(c,t)} $}
\vspace{3ex}

\noindent 
where $W_x^{(c,t)}$ denotes the waveform of cluster $x$ on channel $c$ at time $t$. Since the objective is to align clusters across segments rather than assess whether they should be merged, capturing fine-grained details is unnecessary. Instead, retaining only the most salient waveform features ensures accurate alignment while minimizing computational overhead.

\subsubsection{Electrode Drift Simulation}
\label{subsec:simulation}

\noindent 
To align clusters across segment boundaries, SpikeSift models drift as axial motion along the probe---an assumption widely adopted in drift correction methods \cite{kilosort,drift_correction}. Lateral displacements, which would require substantial tissue disruption, are highly constrained in practice \cite{neuropixels2}, making axial drift the dominant factor affecting signal stability in extracellular recordings.

When drift does not align precisely with the inter-electrode spacing, recorded signals no longer correspond exactly to electrode positions, necessitating interpolation to estimate amplitude values at unmeasured locations. As illustrated in Figures~\ref{fig:phase3}d and~\ref{fig:phase3}e, SpikeSift employs linear interpolation to approximate these values efficiently, as high-precision waveform details are unnecessary for establishing correspondences between clusters. However, when neurons drift beyond the probe’s recorded region, interpolation is no longer feasible, and the nearest available channel is used instead to avoid arbitrary extrapolation.

To balance computational efficiency with the ability to accommodate larger displacements, the maximum tested shift, $D_{\max}$, serves as a tunable parameter, restricting the search space while maintaining sufficient flexibility for alignment. Within this range, discrete drift shifts are evaluated at fixed $5 \mu$m steps, ensuring precise yet computationally efficient alignment. Because SpikeSift guarantees that each neuron is consistently represented by a single cluster within each segment, small interpolation errors in amplitude vectors have negligible impact on final alignment, eliminating the need for finer step sizes.

\subsubsection{Optimal Cluster Matching}
\label{subsec:assignment}

\noindent 
The final step involves establishing the optimal correspondence between clusters in consecutive segments. This is formulated as a generalized linear assignment problem \cite{linear_assignment}, ensuring that neuron assignments remain consistent while minimizing discrepancies.

For each candidate displacement $\delta$, the Euclidean distance between clusters in adjacent segments is computed as:

\vspace{2ex}
\centerline{$ d_{\delta}(x,y)= \|A_x' - A_y'\| $}
\vspace{2ex}

\noindent 
where $A_x'$ and $A_y'$ represent the amplitude vectors of clusters $x$ and $y$ after applying simulated displacements of $\delta/2$ and $-\delta/2$, respectively. This symmetric displacement model ensures balanced alignment, reducing the risk of neurons shifting entirely out of the recorded region.

The total alignment cost for each candidate shift $\delta$ is then computed as:

\vspace{3ex}
\centerline{$ C_{\delta}=\sum_{(x,y)\in M_{\delta}} d_{\delta}(x,y) $}
\vspace{3ex}

\noindent 
where $M_{\delta}$ is the one-to-one cluster matching that minimizes $C_{\delta}$. The optimal shift $\delta^*$  then selected by minimizing this cost:

\vspace{3ex}
\centerline{$ \delta^*=\arg\min_{\delta}C_{\delta} $}
\vspace{3ex}

\noindent 
As illustrated in Figure~\ref{fig:phase3}f, this procedure establishes the final alignment, ensuring that clusters in Segment $A$ are correctly paired with their corresponding clusters in Segment $B$. While waveform-level precision is not exact, the alignment remains sufficiently accurate for maintaining neuronal identity across segments, highlighting that exact waveform matching is not a prerequisite for reliable spike sorting.

\section{Datasets}
\label{datasets}

\noindent 
To rigorously evaluate SpikeSift, we employed a dual-validation framework integrating intracellularly validated recordings (Section~\ref{sec:intracellular}) and biophysically realistic simulations (Section~\ref{sec:simulations}). Intracellular recordings provide highly reliable ground truth for specific neurons, confirming whether detected spikes correspond to actual neuronal activity. However, they offer no insight into whether additional detected clusters reflect genuine units or spurious detections. To address this limitation, we complement intracellular recordings with biophysically realistic simulations, which provide exact spike times for all neurons, enabling a complete assessment of both detection and clustering accuracy. Together, these complementary approaches offer a comprehensive evaluation: intracellular data test real-world detection performance, while simulations verify the validity of all detected clusters.

\subsection{Paired Intracellular Recordings}
\label{sec:intracellular}

\noindent 
We evaluated SpikeSift using 12 extracellular recordings \cite{experimental_dataset}, each paired with an intracellular recording that provided precisely identified firing events. Originally introduced for the validation of SpyKING CIRCUS \cite{circus}, these recordings serve as a well-established benchmark for assessing spike sorting accuracy. Extracellular signals were acquired using a high-density microelectrode array consisting of 252 active channels arranged in a $16 \times 16$ grid with an inter-electrode spacing of $30 \mu$m.

The corpus originally comprised 19 paired recordings. Meaningful quantitative evaluation requires an intracellular trace whose spike events can be identified without ambiguity; otherwise, accuracy scores for any sorter would be unreliable. We therefore inspected the histogram of positive intracellular peaks in each file (16-bit integer samples). A recording was accepted when at least one unused integer value---an empty bin---separated the dense noise mode from a sparser band of spike amplitudes. Eleven of the twelve retained files showed a gap $>$$20 \mu V$; the twelfth exhibited a shorter $2 \mu V$ gap yet still displayed a clear valley. Gap size bore no relation to sorter accuracy---indeed, the file with the smallest gap was processed flawlessly by every method, whereas the first file that challenged all sorters contained a $170 \mu V$ gap. The seven recordings that lacked any empty bin were excluded because noise and spike distributions overlapped continuously, making ground-truth labels unreliable. The twelve recordings that met the gap criterion constitute the benchmark summarised in Table \ref{tab:intracellular}.

\begin{table}
\setlength{\tabcolsep}{0.02\textwidth}
\caption{Summary of paired intracellular recordings.}
\begin{tabular}{@{}cccr@{}}
\br
ID & Name & Seconds & Spikes \\
\mr
\phantom{0}1   & 20160415\_patch2 & 300  &   3514  \\
\phantom{0}2   & 20160426\_patch2 & 202  &    879  \\
\phantom{0}3   & 20160426\_patch3 & 180  &   1691  \\
\phantom{0}4   & 20170621\_patch1 & 300  &   4998  \\
\phantom{0}5   & 20170622\_patch1 & 300  &   4541  \\
\phantom{0}6   & 20170623\_patch1 & 300  &    737  \\
\phantom{0}7   & 20170630\_patch1 & 300  &   2385  \\
\phantom{0}8   & 20170713\_patch1 & 300  &   6557  \\
\phantom{0}9   & 20170725\_patch1 & 300  &    380  \\
10             & 20170726\_patch1 & 300  &   2413  \\
11             & 20170728\_patch2 & 300  &   4748  \\
12             & 20170803\_patch1 & 300  &   7639  \\
\br
\end{tabular}
\label{tab:intracellular}
\end{table}

\subsection{Biophysically Realistic Simulations}
\label{sec:simulations}

\noindent 
To complement intracellular validation, we employed simulated recordings generated using MEArec \cite{mearec}, a widely used framework \cite{removing_noise,drift_correction} that synthesizes extracellular activity based on biophysically detailed neuron models from the Neocortical Microcircuit Collaboration Portal \cite{neocortical_portal}. These simulations provide a controlled yet biologically plausible environment for evaluating spike sorting performance across diverse recording conditions, including scenarios with electrode drift.

To construct a biologically diverse neuronal population, we randomly positioned 384 neurons within $100  \mu$m of the recording probe \cite{spike_sorting_overview}. Although low-amplitude neurons cannot be reliably sorted, their presence preserves the natural complexity of extracellular activity, ensuring that the dataset closely resembles in vivo conditions. However, to balance realism with computational efficiency, we imposed a $10 \mu$V amplitude threshold, corresponding to half the background noise standard deviation of $20 \mu$V \cite{removing_noise,mearec}. This threshold ensures that all 384 neurons meaningfully contribute to the recorded extracellular potential

To further enhance realism, we modeled the recording probe after the Neuropixels 2.0 array \cite{neuropixels2}, which consists of 384 densely packed electrodes arranged in two columns with $15 \mu$m intra-column spacing and $32 \mu$m inter-column spacing. Neurons were placed independently, maintaining a minimum separation of $15 \mu$m, consistent with experimental estimates of cortical neuron densities \cite{cell_densities}. This setup ensures that simulated recordings accurately reflect the electrode configurations used in large-scale extracellular experiments. Additionally, neuronal firing rates were sampled from a uniform distribution between 1 and $50 Hz$, capturing the variability observed in cortical networks \cite{firing_rates}, while the simulations were conducted at $20 kHz$ to align with standard extracellular recording protocols without introducing unnecessary computational overhead \cite{spike_sorting_overview}.

To probe SpikeSift’s robustness under controlled yet challenging scenarios, 13 MEArec datasets were generated.  The suite comprised a 2-minute baseline recording with $20 \mu V$ RMS noise; two further 2-minute recordings with elevated noise levels of $30 \mu V$ and $40 \mu V$; a 5-second ultra-short trace to test performance with limited spike counts; three 10-minute recordings exhibiting distinct drift regimes---linear drift at $5 \mu m \cdot min^{-1}$, a non-linear Gaussian-process trajectory spanning $50 \mu m$ peak-to-peak, and abrupt displacements of 20--$25 \mu m$ every two minutes; a set of five 1-minute recordings sweeping linear drift from 5 to $25 \mu m \cdot min^{-1}$; and a 1-minute recording containing thirty intrinsically bursting neurons.  Collectively, these datasets span key nuisance factors---noise amplitude, spike scarcity, drift geometry and magnitude, and non-stationary firing patterns.

\section{Experiments}
\label{experiments}

\noindent
SpikeSift was evaluated against three widely used spike sorting algorithms---Kilosort \cite{kilosort}, Mountainsort \cite{mountainsort}, and SpyKING CIRCUS \cite{circus}---which differ substantially in their strategies for spike detection, clustering, and electrode drift compensation. This diversity enables a robust comparative assessment across methodological paradigms. All algorithms were tested under standardized hardware conditions, using consistent evaluation metrics and carefully tuned parameters.

\subsection{Computational Setup}
\label{sec:setup}

\noindent 
Benchmarks were performed on a standardized system with an Intel i7-4790K CPU, 16 GB RAM, and an NVIDIA GTX 980 GPU. SpikeSift and Mountainsort were restricted to a single CPU core, while Kilosort used GPU acceleration and SpyKING CIRCUS leveraged multithreading, reflecting their native execution models.

\subsection{Performance Evaluation}
\label{sec:evaluation}

\noindent 
Sorting accuracy was evaluated by comparing detected spike times to ground-truth firing events, considering a match when a detected spike occurred within $0.5 ms$ of a ground-truth spike. Performance remained consistent across slight variations in this threshold, demonstrating robustness to minor timing discrepancies.

For experimental recordings, where intracellular measurements provided ground-truth firing times for a single neuron, the detected cluster with the highest match count was selected for evaluation.

For simulated recordings, detected units were assigned to ground-truth neurons using the scoring metric from \cite{kilosort}, defined as:

\vspace{2ex}
\centerline{$ Score = 1 - FP - FN $}
\vspace{2ex}

\noindent 
where false positives ($FP$) correspond to detected spikes without a matching ground-truth event, and false negatives ($FN$) denote missed ground-truth spikes. Clusters scoring above 0.95 were classified as identified, while those scoring below 0.8 were considered spurious. Intermediate scores were left unclassified to avoid imposing arbitrary thresholds.

\subsection{Comparison with Other Algorithms}
\label{sec:comparison}

\noindent 
To ensure a fair comparison, the latest version of each algorithm was used, with default parameters modified only where necessary to enhance performance. These adjustments ensured that observed differences reflected methodological distinctions rather than implementation discrepancies.

Kilosort 4 is a GPU-accelerated spike sorting algorithm that combines template matching with graph-based clustering. It employs a continuous drift correction mechanism that estimates neuron displacement over time and interpolates waveforms to compensate for motion. Initial tests showed that the default detection thresholds led to excessive spurious detections. To mitigate this, the initial detection threshold was increased to 12, while the secondary threshold was set to 8, balancing sensitivity and clustering accuracy.

Mountainsort 5 segments recordings into discrete time blocks, performing clustering independently within each segment before merging clusters. It employs ISO-SPLIT \cite{iso_split}, a clustering algorithm that recursively partitions data through unimodality tests, allowing it to adapt to complex waveform distributions without assuming predefined cluster shapes. To enhance drift adaptation, the block size was reduced from five minutes to one minute, allowing for finer adjustments. Initial trials showed that default PCA settings retained excessive dimensions, introducing noise rather than highlighting informative waveform differences. To improve feature extraction, the number of principal components per channel was reduced to 1, while nPCA was set to 20, enhancing cluster separability while minimizing unnecessary complexity.

SpyKING CIRCUS is a multi-core spike sorting algorithm that combines density-based clustering with template matching. It iteratively refines spike assignments by reconstructing extracellular signals as sums of individual waveform templates, allowing it to resolve overlapping spikes. Initial trials revealed a tendency for over-splitting, prompting an increase in the clustering sensitivity parameter from 3 to 10, which reduced fragmentation while maintaining unit separability.

\subsection{SpikeSift Parameter Configuration}
\label{sec:parameters}

\noindent 
Unlike conventional spike sorting pipelines that require extensive parameter tuning, SpikeSift is designed with interpretable parameters that provide direct control over fundamental sorting trade-offs. This flexibility allows parameter optimization based on specific priorities, such as detecting low-amplitude neurons, reducing false positives, or maximizing processing speed.

The spike detection sensitivity parameter was set to $\kappa = 10$, controlling the trade-off between spike detection and sorting reliability. Lower values increase sensitivity, capturing more spikes at the cost of potential false positives, while higher values restrict detection to the most prominent spikes, enhancing clustering accuracy.

The cluster merging threshold was set to $\lambda = 0.4$, balancing neuron identity preservation with the risk of merging distinct neurons. Since neuronal waveforms naturally vary due to noise and firing conditions, $\lambda$ determines how much waveform deviation is tolerated before two clusters are considered distinct. A higher $\lambda$ promotes cluster merging, potentially conflating distinct neurons, whereas a lower $\lambda$ enforces stricter separation, reducing the likelihood of merging at the expense of an increased risk of over-splitting, where natural waveform variability causes spikes from the same neuron to be divided into multiple clusters. Sensitivity analysis in Section~\ref{sec:sensitivity} confirmed that sorting performance remains stable across reasonable variations in $\lambda$, ensuring robustness to moderate waveform fluctuations.

The minimum required cluster size was set to $N_{\min} = 5$, balancing the detection of low-firing neurons with clustering reliability. A lower value allows the identification of infrequently firing neurons, while a higher value reduces the likelihood of forming spurious clusters, ensuring that only neuronal units with sufficient spike counts are retained.

The minimum segment duration was set to $L_{\min} = 10$ seconds, consistent with the common assumption---even in global drift correction methods---that recording conditions remain stationary over short time intervals \cite{drift_correction}. This prevents excessive segmentation while allowing adaptation to slow drift dynamics. Users working with particularly stable recordings may opt for longer segments, while those dealing with frequent drift fluctuations may benefit from reducing $L_{\min}$.

The maximum electrode displacement accounted for was set to $D_{\max} = 30 \mu$m, encompassing the range of most experimentally observed drift magnitudes \cite{spike_sorting_overview}. This choice balances computational efficiency and robustness, as lower values accelerate processing, while higher values accommodate larger displacements, ensuring stable clustering even in recordings with significant electrode motion.

\begin{figure*}[t]
    \centering
    \includegraphics[width=0.9\textwidth,height=0.45\textwidth]{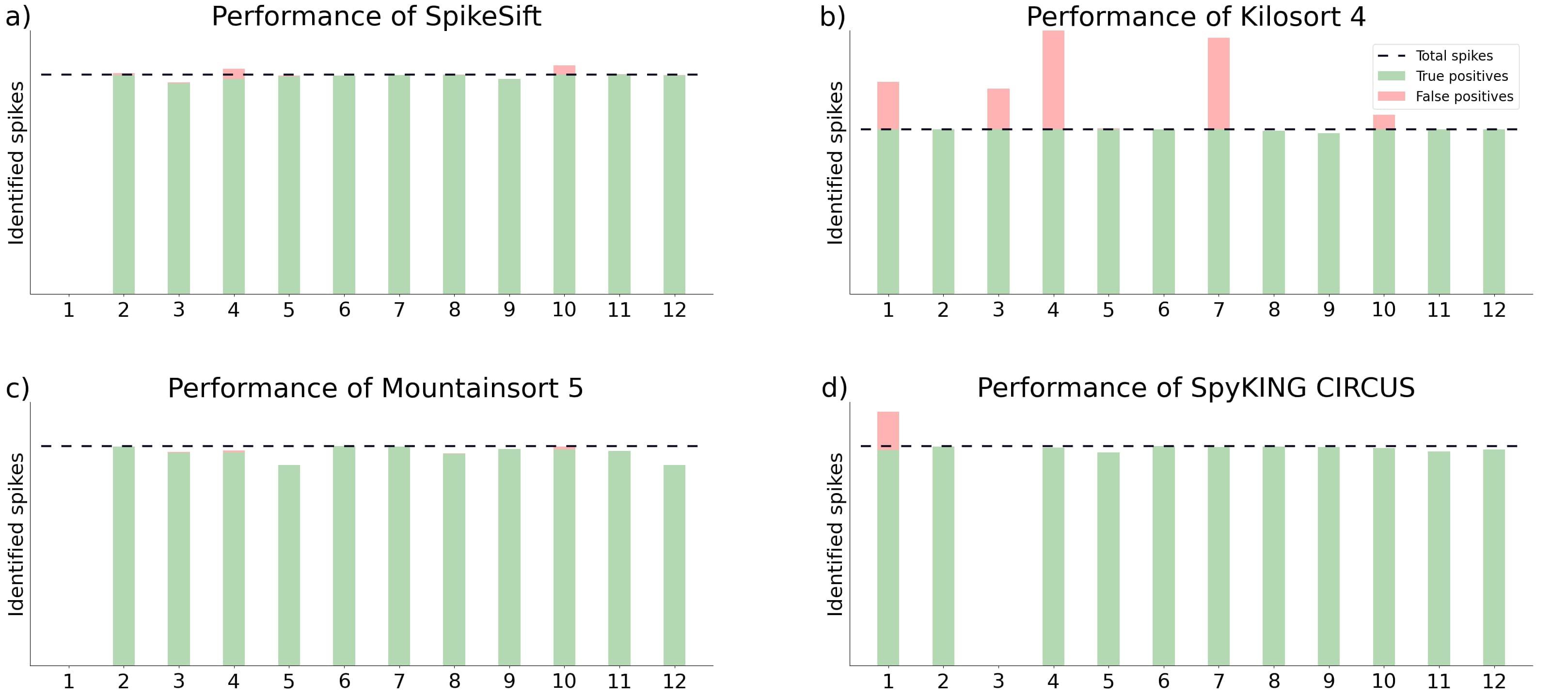}
    \caption{Sorting accuracy on intracellularly validated recordings for each spike sorting algorithm, with each subfigure corresponding to a different method. Bars represent individual recordings, with the horizontal dashed line indicating the total number of ground-truth spikes. The green portion of each bar represents correctly identified spikes, while the red portion denotes detected spikes that do not match the ground-truth validation. Some bars are missing because the corresponding algorithm failed to detect any of the intracellularly validated spikes in those recordings.}
    \label{fig:intracellular}
\end{figure*}

\begin{figure*}[t]
    \centering
    \includegraphics[width=0.9\textwidth,height=0.64\textwidth]{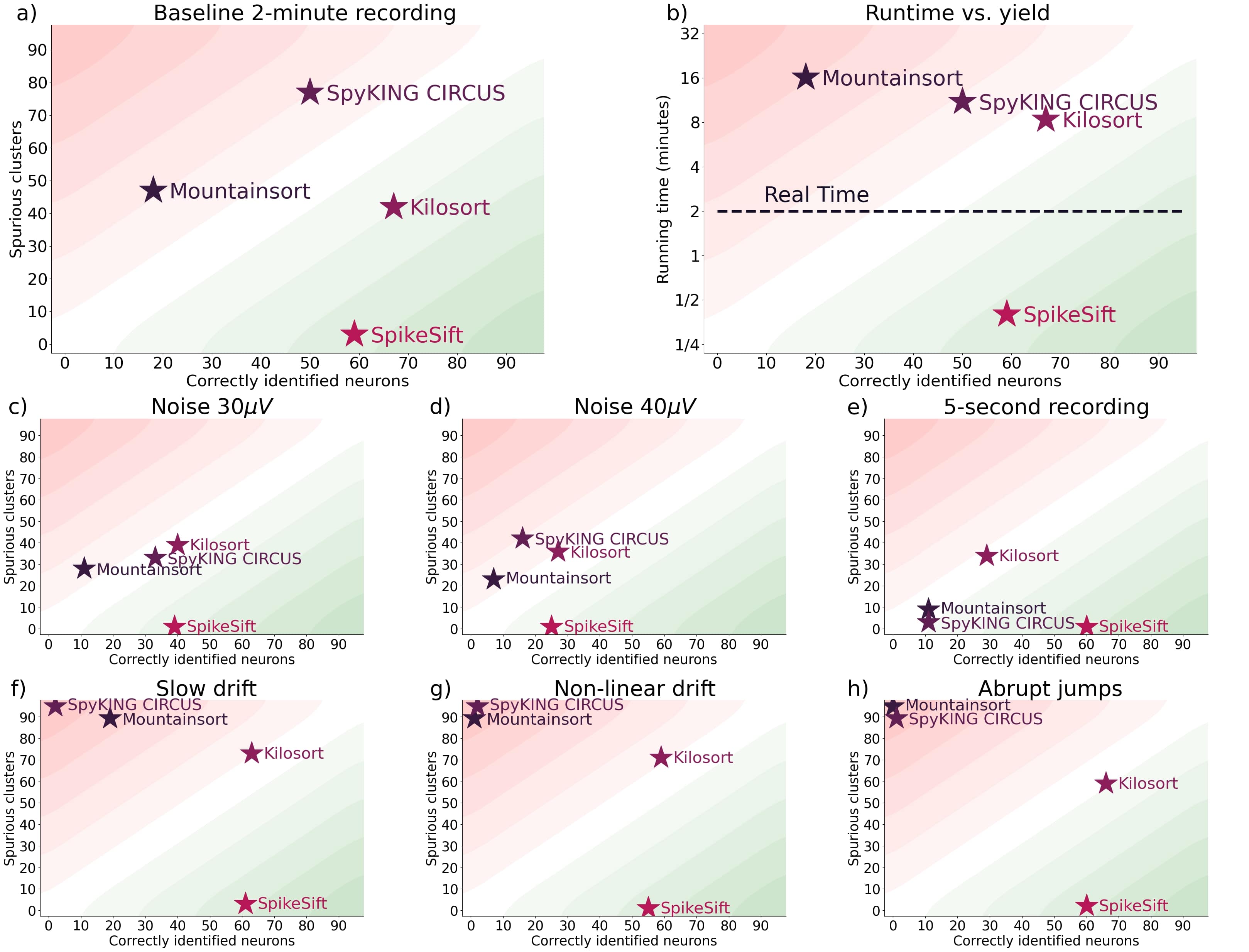}
    \caption{Sorting performance under eight controlled conditions. Each bar shows correctly identified neurons (horizontal axis) against either spurious clusters (panels a, c--h; vertical axis) or runtime in minutes (panel b). 
    (a) baseline 2-minute recording;
    (b) computational efficiency across methods;
    (c--d) elevated-noise recordings ($30 \mu V$, $40 \mu V$ RMS); 
    (e) ultra-short 5-second recording; 
    (f) slow linear drift ($5 \mu m \cdot min^{-1}$);
    (g) non-linear Gaussian-process drift spanning $50 \mu m$;
    (h) abrupt 20--$25 \mu m$ jumps every 2 minutes.
    }
    \label{fig:mearec}
\end{figure*}

\section{Results}
\label{results}

\noindent
SpikeSift was evaluated alongside Kilosort, Mountainsort, and SpyKING CIRCUS using both intracellularly validated and biophysically realistic simulated recordings. Evaluations focused on spike sorting accuracy, computational efficiency, and robustness under challenging conditions. Across all benchmarks, SpikeSift achieved comparable or superior accuracy to established methods while offering a substantial improvement in runtime efficiency. In addition, SpikeSift maintained stable performance across a broad range of parameter settings, confirming its adaptability to diverse experimental scenarios.

\subsection{Sorting Performance}
\label{sec:performance}

\noindent 
Figure~\ref{fig:intracellular} summarizes sorting accuracy on experimentally recorded datasets. Each panel corresponds to a different algorithm, with bars representing individual recordings and the dashed line denoting the total number of ground-truth spikes. All methods successfully identified most ground-truth spikes, but their computational demands varied considerably.

Running on a single CPU core, SpikeSift consistently achieved a 10-fold speed advantage over Kilosort, Mountainsort, and SpyKING CIRCUS---despite their use of GPU acceleration or multi-threaded execution. For instance, Recording 2 from Table~\ref{tab:intracellular} was processed in just 13 seconds using SpikeSift, compared to 200 seconds for Kilosort and over 300 seconds for both Mountainsort and SpyKING CIRCUS. These performance differences persisted in simulated recordings, as shown in Figure~\ref{fig:mearec}b, where SpikeSift maintained a 20--$40 \times$ speed advantage despite the additional complexity associated with the increased channel count. Across all datasets SpikeSift used less than 3 GB of RAM and wrote no temporary files, whereas competing pipelines created full on-disk duplicates of the filtered data and, for the 10-minute simulations, generated an additional $\approx$ 18 GB of intermediates. SpikeSift’s resource footprint is therefore essentially flat with recording length, while the alternatives scale at least linearly.

While computational efficiency is essential for large-scale applications, reliability of sorting results is equally critical---especially in experimental datasets where only a subset of neurons can be validated. In these recordings, SpikeSift successfully identified all ground-truth neurons except one. Notably, this neuron can be recovered by slightly lowering the detection threshold to $\kappa \leq 8$, though---as with other algorithms---this increases the number of additional, biologically unverified clusters.

Simulated recordings, which provide exhaustive ground truth for all neurons, further clarify this trade-off. As shown in Figure~\ref{fig:mearec}a, SpikeSift recovered 59 neurons with only 3 spurious clusters at the default detection threshold. As summarized in Table~\ref{tab:kappa}, reducing $\kappa$ to 7 increases sensitivity to match Kilosort, recovering 67 neurons, while still producing significantly fewer spurious clusters (21 vs. 42). This result illustrates how a single tunable parameter enables users to control the balance between sensitivity and specificity, offering a principled and transparent approach to adapting spike sorting behavior across diverse experimental demands.

Raising background noise from $20 \mu V$ to $30 \mu V$ and $40 \mu V$ reduced unit yield for every sorter, yet their rank order did not change. At $40 \mu V$ SpikeSift still recovered 25 ground-truth neurons while producing a single false cluster. Kilosort returned 27 true units but 36 spurious clusters, and SpyKING CIRCUS and Mountainsort lagged further, at 16~/~5 and 7~/~26 true~/~false units, respectively (Figures~\ref{fig:mearec}c and~\ref{fig:mearec}d.).

\begin{table*} 
\setlength{\tabcolsep}{0.0580\textwidth}
\caption{Effect of detection threshold $\kappa$ on SpikeSift performance (2-minute baseline recording).}
\begin{tabular}{@{}cccccc@{}} 
\br 
$\kappa$ & 7 & 9 & 10 & 11 & 13 \\
\mr
Correctly identified neurons & 67 & 61 & 59 & 55 & 50 \\
Spurious clusters & 21 & 10 & \phantom{0}3 & \phantom{0}1 & \phantom{0}0 \\
\br 
\end{tabular} 
\label{tab:kappa} 
\end{table*}

A one-minute MEArec recording containing thirty intrinsically bursting neurons challenged each pipeline with strongly patterned firing. With its default threshold ($\kappa = 10$) SpikeSift isolated 24 genuine units and produced no false clusters; Kilosort, SpyKING CIRCUS and Mountainsort reported 27~/~6, 16~/~5 and 7~/~26 true~/~false units, respectively. Lowering $\kappa$ to 9 elevated SpikeSift’s recall to 28 while preserving its zero-false-cluster record, matching Kilosort’s sensitivity but at markedly higher precision---a direct extension of the sensitivity--specificity trade-off shown in Table \ref{tab:kappa}.

\subsection{Robustness to Electrode Drift}
\label{sec:drift}

\noindent 
Sorting performance diverged most clearly in the five-second recording (Figure~\ref{fig:mearec}e), where limited spike counts posed a challenge for conventional methods. While accuracy declined sharply for other algorithms, SpikeSift maintained stable performance, demonstrating that its iterative detect-and-subtract framework remains effective even under sparse data conditions. This result highlights how SpikeSift overcomes the limitations that have traditionally hindered segmentation-based strategies in conventional pipelines.

This advantage becomes even more evident in the presence of electrode drift, as shown in Figures~\ref{fig:mearec}f, \ref{fig:mearec}g  and~\ref{fig:mearec}h. Mountainsort and SpyKING CIRCUS struggled to maintain consistent neuron identities over time due to limited or absent drift correction. SpyKING CIRCUS lacks any compensation mechanism, while Mountainsort processes data in fixed-duration blocks and applies a merging procedure that does not explicitly account for drift-induced shifts in waveforms.

Even in Kilosort, which incorporates drift correction, small errors in trajectory estimation propagated through the pipeline, disrupting template matching and destabilizing cluster assignments. In contrast, SpikeSift maintained reliable performance across both gradual and abrupt drift scenarios. Its segmentation-based framework enables local adaptation without relying on accurate drift trajectories, providing a robust and fault-tolerant solution for non-stationary recordings.

\begin{table*} 
\setlength{\tabcolsep}{0.0523\textwidth}
\caption{Effect of merging threshold $\lambda$ on SpikeSift performance (2-minute baseline recording).}
\begin{tabular}{@{}cccccc@{}} 
\br 
$\lambda$ & 0.3 & 0.35 & 0.4 & 0.45 & 0.5 \\
\mr
Correctly identified neurons & 58 & 58 & 59 & 59 & 56 \\
Spurious clusters & \phantom{0}5 & \phantom{0}6 & \phantom{0}3 & \phantom{0}5 & \phantom{0}6 \\
\br 
\end{tabular} 
\label{tab:lambda} 
\end{table*}

\subsection{Sensitivity Analysis}
\label{sec:sensitivity}

\noindent 
SpikeSift offers a flexible trade-off between sensitivity and specificity through a single tunable parameter, $\kappa$, which governs the spike detection threshold. Unlike conventional pipelines---where conservative thresholds may fragment neuronal units by capturing only a subset of spikes---SpikeSift preserves complete neuronal representations even when operating at reduced sensitivity.

As shown in Table~\ref{tab:kappa}, lowering $\kappa$ to 7 enables SpikeSift to match Kilosort’s sensitivity, recovering 67 neurons, while producing only half as many spurious clusters (21 vs. 42). Conversely, increasing $\kappa$ to 13 eliminates all spurious detections while still recovering 50 valid neurons---demonstrating a level of specificity unmatched by competing methods. These low false-positive counts show that the default pairing of a conservative detection threshold and a strict inter-cluster similarity test biases SpikeSift toward well-isolated single-unit clusters rather than broader multi-unit aggregates.

SpikeSift also demonstrates strong resilience to variations in other core parameters. As shown in Table~\ref{tab:lambda}, it maintains stable performance across a broad range of merging thresholds $\lambda$, consistently recovering 56--59 neurons with minimal impact on precision. Likewise, Table~\ref{tab:Lmin} shows that even under high drift rates ($25,\mu$m/min), SpikeSift detects 54 neurons with only five spurious clusters---confirming that adaptive segmentation remains robust in highly non-stationary environments.

These results underscore the flexibility of $L_{\min}$, the minimum segment duration. The default value of 10 seconds suffices to track fast drift, while the five-second benchmark in Figure~\ref{fig:mearec}e illustrates that shorter segments remain effective when required.

Robustness also extends to clustering criteria. Lowering the minimum cluster size $N_{\min}$ to 2 introduces only four additional spurious units---indicating that even low-firing or sparsely active neurons can be detected without substantial degradation in precision.

Finally, expanding the drift correction range $D_{\max}$ has negligible computational cost: across-segment merging accounts for less than 1\% of total runtime, making it safe to increase the alignment range when necessary.

Three recording contingencies may nonetheless warrant deliberate departures from the default settings. Substantial, short-interval drift ($>$$25 \mu m$) is accommodated by widening the drift-alignment radius $D_{\max}$ and shortening the segment length $L_{\min}$ so that each segment remains approximately stationary. Recordings dominated by rarely firing neurons benefit from longer segments and a lower cluster-size threshold ($L_{\min} \uparrow$,$N_{\min} \downarrow$); a modest increase in the detection threshold $\kappa$ then suppresses noise that would otherwise contaminate sparsely populated clusters. Conversely, when low-amplitude spikes are the limiting factor, lowering $\kappa$ improves sensitivity, while a compensatory rise in $N_{\min}$ curbs the formation of noise-driven clusters.

Together, these findings confirm that SpikeSift delivers reliable spike sorting across a broad range of parameter settings---maintaining high accuracy and interpretability with minimal need for manual tuning, even under challenging recording conditions.

\begin{table*} 
\setlength{\tabcolsep}{0.0580\textwidth}
\caption{SpikeSift performance under increasing drift (five 1-minute recordings, 5--$25\,\mu$m/min).}
\begin{tabular}{@{}cccccc@{}} 
\br 
Drift ($\mu$m/min) & 5 & 10 & 15 & 20 & 25 \\
\mr
Correctly identified neurons & 57 & 59 & 57 & 56 & 54 \\
Spurious clusters & \phantom{0}1 & \phantom{0}1 & \phantom{0}3 & \phantom{0}2 & \phantom{0}5 \\
\br 
\end{tabular} 
\label{tab:Lmin} 
\end{table*}

\section{Discussion}
\label{discussion}

\noindent Despite its strong performance across diverse recording conditions, SpikeSift’s current implementation presents several opportunities for refinement and future development.

One area for improvement lies in the detection of low-firing or transient neurons that do not appear consistently across segments. Although SpikeSift’s current pipeline is optimized for stable, recurring activity, such cases can be addressed by first sorting a short recording segment spanning a few minutes (which will still be internally segmented), and then aligning subsequent segments to it using SpikeSift’s existing merging mechanism. This approach enables the recovery of units even when their activity is intermittent and incurs less than 1\% additional runtime, making it practical within existing workflows.

A second limitation is the absence of built-in support for parallel execution. Although SpikeSift is already highly optimized for single-core performance---outperforming GPU-accelerated methods like Kilosort---the current implementation does not internally parallelize across segments. Nevertheless, because segments are processed independently and merging incurs negligible overhead, the architecture is inherently well-suited to parallel or distributed execution. This design allows additional computational resources to be leveraged with minimal modification, enabling scalable deployment on multi-core systems or cluster environments as needed.

Finally, while SpikeSift supports progressive sorting of streaming or growing recordings, it introduces a minimal latency governed by the minimum segment duration parameter $L_{\min}$. Reducing this latency to achieve true millisecond-level spike sorting under dynamic conditions---particularly in the presence of abrupt drift---remains an open challenge in the field. Nevertheless, SpikeSift already processes Neuropixels-scale data more than five times faster than real time on a single CPU core, providing a strong foundation for future real-time or closed-loop applications.

Looking ahead, SpikeSift is distributed as a lightweight Python package designed for scripted, reproducible workflows \cite{spikesift_zenodo}.  A graphical user interface that displays segment boundaries, cluster assignments, and merge operations---while allowing manual overrides---is a high-priority direction for future development.  In the meantime, users can inject domain expertise through the existing programmatic hooks: a sorted recording can be divided with \texttt{split\_into\_segments}, rerun on selected portions via \texttt{perform\_spike\_sorting} with modified parameters, and curated or reconciled with \texttt{merge\_recordings} and \texttt{map\_clusters}.  These tools allow experts to inspect, refine, and combine clusters directly within a scripted environment, offering a transparent and reproducible means of incorporating manual oversight.

\section{Conclusion}
\label{conclusion}

\noindent 
SpikeSift introduces a fast, modular, and drift-resilient framework for spike sorting, designed to meet the growing demands of high-density electrophysiology. By combining adaptive segmentation with an iterative detect-and-subtract strategy, it eliminates the need for global drift trajectory estimation while achieving high accuracy, strong cluster consistency, and exceptionally low computational cost. SpikeSift maintains stable performance even under challenging conditions---including abrupt drift, low firing rates, and short-duration recordings---where conventional pipelines often fail.

Extensive evaluations on both intracellularly validated and realistically simulated datasets demonstrate that SpikeSift consistently matches the sensitivity of established methods like Kilosort, while producing significantly fewer spurious clusters. At the same time, it delivers a dramatic gain in computational efficiency, processing large-scale datasets more than ten times faster than GPU-based alternatives, despite running on a single CPU core. This combination of speed and accuracy makes it particularly well-suited for time-sensitive or resource-constrained applications, without relying on specialized hardware.

SpikeSift’s modular architecture supports both progressive and parallel workflows, enabling flexible sorting and merging of recordings---including split files, streaming pipelines, or segments excluded post hoc due to noise or artifacts. These capabilities are achieved with negligible computational overhead and without compromising accuracy or interpretability. Crucially, SpikeSift maintains stable performance across a broad range of parameter settings, minimizing the need for manual tuning and promoting reproducible results.

As neural recordings continue to expand in size, density, and duration, the need for efficient and interpretable spike sorting becomes increasingly critical. SpikeSift rises to this challenge, offering a principled and practical solution for modern neuroscience---whether in high-throughput data pipelines, closed-loop experimental systems, or scalable brain--computer interface applications. Future work will focus on further reducing latency, improving support for low-firing and transient units, and extending the framework toward fully online spike sorting under non-stationary conditions.

\section*{Acknowledgments}
This work was partially supported by the project “DeepTector”, funded by the Special Account for Research Funds of the Aristotle University of Thessaloniki (project no. 10716).

\section*{Code Availability}

\noindent SpikeSift is available on Zenodo under the MIT license \cite{spikesift_zenodo}, with installation instructions, usage examples, and documentation provided at \url{https://spikesift.readthedocs.io}.

\end{document}